\documentclass[12pt,english]{article}
\usepackage{graphicx,array}
\usepackage{cite}
\usepackage{color}
\usepackage{latexsym}
\usepackage{amsthm}
\usepackage{amsmath}
\usepackage[titletoc]{appendix}
\usepackage{enumitem}
\usepackage{amssymb}
\usepackage{hyperref}
\usepackage[hang,flushmargin]{footmisc} 
\usepackage{stmaryrd}
\usepackage{mathtools}

\allowdisplaybreaks[1]

\def\simgt{\mathrel{\lower2.5pt\vbox{\lineskip=0pt\baselineskip=0pt
           \hbox{$>$}\hbox{$\sim$}}}}
\def\simlt{\mathrel{\lower2.5pt\vbox{\lineskip=0pt\baselineskip=0pt
           \hbox{$<$}\hbox{$\sim$}}}}

\newcommand{\be}{\begin{equation}}
\newcommand{\ee}{\end{equation}}
\newcommand{\bea}{\begin{eqnarray}}
\newcommand{\eea}{\end{eqnarray}}
\newcommand{\Eq}[1]{Eq.~(\ref{#1})}
\newcommand{\Eqs}[2]{Eqs.~(\ref{#1}) and (\ref{#2})}
\newcommand{\Sec}[1]{Sec.~\ref{#1}}

\newcommand{\App}[1]{App.~\ref{#1}}

\newcommand{\Ref}[1]{Ref.~\cite{#1}}

\newcommand{\Disc}[1]{\mathrm{Disc}[#1]}
\newcommand{\Msym}{\mathcal{M}_\mathrm{sym}}
\newcommand{\Tr}{\mathrm{Tr}}

\newcommand{\lambdaGB}{\lambda}

\renewcommand{\Msym}{M}
\newcommand{\Mforw}{M}

\newcommand*\oline[1]{%
  \vbox{%
    \hrule height 0.5pt
    \kern0.68ex
    \hbox{%
      \kern-0.1em
      \ifmmode#1\else\ensuremath{#1}\fi
      \kern-0.1em
    }
  }
}

\definecolor{nicered}{rgb}{0.7,0.1,0.1}
\definecolor{nicegreen}{rgb}{0.1,0.5,0.1}
\hypersetup{
 colorlinks,citecolor=black,linkcolor=black,urlcolor=black}

\begin{document}
\interfootnotelinepenalty=10000
\baselineskip=18pt
\hfill CALT-TH-2015-044 

\hfill Saclay-t15/161

\vspace{1.5cm}
\thispagestyle{empty}
\begin{center}
{\LARGE\bf
Quantum Gravity Constraints \\ \medskip
 from Unitarity and Analyticity
}\\

\bigskip\vspace{1.5cm}{
{\large Brando Bellazzini,$^1$ Clifford Cheung,$^2$ and Grant N. Remmen$^2$}
} \\[7mm]
{\it $^1$Institut de Physique Th\'eorique, Universit\'e Paris Saclay, CEA, CNRS, \\[-1mm]
F-91191 Gif-sur-Yvette, France \\
$^2$Walter Burke Institute for Theoretical Physics, 
\\[-1mm] California Institute of Technology, Pasadena, CA 91125, USA}
\let\thefootnote\relax\footnote{e-mail:
\url{brando.bellazzini@cea.fr}, \url{clifford.cheung@caltech.edu}, \url{gremmen@theory.caltech.edu}} \\
 \end{center}
\bigskip
\centerline{\large\bf Abstract}

\begin{quote}\small
We derive rigorous bounds on corrections to Einstein gravity using unitarity and analyticity of graviton scattering amplitudes. In $D\geq 4$ spacetime dimensions, these consistency conditions mandate positive coefficients for certain quartic curvature operators.  We systematically enumerate all such positivity bounds in $D=4$ and $D=5$ before extending to $D\geq 6$.  Afterwards, we derive positivity bounds for supersymmetric operators and verify that all of our constraints are satisfied by weakly-coupled string theories.  Among quadratic curvature operators, we find that the Gauss-Bonnet term in $D\geq5 $ is inconsistent unless new degrees of freedom enter at the natural cutoff scale defined by the effective theory. Our bounds apply to perturbative ultraviolet completions of gravity.
\end{quote}

\setcounter{footnote}{0}

\newpage
\tableofcontents

\newpage

\section{Introduction}

Low-energy effective theory describes quanta interacting indirectly through kinematically inaccessible states.  The dynamics are characterized by an effective action that typically includes all interactions permitted by symmetry with couplings of order unity.  However, in certain cases, self-consistency at long distances imposes non-trivial constraints on the coefficients of effective operators.  This is famously true in the theory of pions, where the operator coefficients of certain higher-derivative operators are required to be strictly positive to ensure causal  particle propagation together with unitarity and analyticity of scattering amplitudes at complex momenta \cite{IRUV,Brando,Rothstein,Manohar:2008tc,Pham:1985cr}. 

As low-energy criteria, causality, unitarity, and analyticity impose  constraints that are independent of the detailed ultraviolet dynamics.  Consequently, these consistency conditions offer special utility in the context of quantum gravity, where the ultraviolet completion is not known with certainty.
 For instance, such bounds have been derived for the effective theory of gravitons and photons \cite{IRWGC}, where consistency necessitates large charge-to-mass ratios precisely of the form of the weak gravity conjecture \cite{WGC}.  

Notably, a proper diagnosis of causality violation in curved spacetime is subtle since particle propagation can be locally superluminal even in healthy theories.  For example, it has long been known that photons with certain polarizations travel superluminally in the vicinity of a black hole in the effective theory of photons and gravitons describing our actual universe \cite{Drummond}.   Instead, a more global measure of causality, {\it e.g.}, the existence of closed timelike curves, is necessary to establish a true pathology.
On the other hand, unitarity and analyticity offer alternative criteria that are mathematically rigorous and applicable in the flat-space limit.

In this paper, we systematically derive new constraints on curvature corrections in gravity from unitarity and analyticity.
The graviton effective theory is described by the action\footnote{We work in mostly $+$ signature and write $\kappa=\sqrt{8\pi G}$, $R_{\mu\nu} = R^\rho_{\;\;\mu\rho\nu}$, and $R^\mu_{\;\;\nu\rho\sigma} = \partial_\rho \Gamma^\mu_{\;\;\nu\sigma}+\cdots$.}
\be 
S = \int {\rm d}^Dx \; \sqrt{-g} \; \sum_{n=1}^\infty {\cal L}_n ,
\ee
where ${\cal L}_n $ are contributions to the action entering at order $2n$ in the derivative expansion and
\be 
\mathcal{L}_1 = \frac{R}{2\kappa^2}   \qquad \textrm{and} \qquad
\mathcal{L}_2 = \lambdaGB (R_{\mu\nu\rho\sigma}R^{\mu\nu\rho\sigma} - 4 R_{\mu\nu} R^{\mu\nu} + R^2)
\ee
are the Einstein-Hilbert and Gauss-Bonnet terms.  We assume the Gauss-Bonnet form for ${\cal L}_2$ throughout since this is the unique ghost-free quadratic curvature invariant \cite{Zwiebach,Zumino:1985dp} in $D$ dimensions.  Moreover, we restrict our analysis to effective theories in which ${\cal L}_4$ takes the form
\be 
\mathcal{L}_4 =  \sum_{i=1}^7 c_i {\cal O}_i ,\label{eq:L4}
\ee 
expressed in terms of the minimal basis of quartic Riemann operators\footnote{Applying leading-order equations of motion to ${\cal L}_n$ is equivalent to a field definition modulo new terms generated in ${\cal L}_{n+1}$.  Repeating this procedure at each order, we can freely impose $R = R_{\mu\nu}=0$ in a pure gravity theory \cite{Georgi:1991ch}.} in \Ref{Invariants},
\be 
\begin{aligned}
&{\cal O}_1 = R^{\mu\nu\rho\sigma}R_{\mu\nu\rho\sigma}R^{\alpha\beta\gamma\delta}R_{\alpha\beta\gamma\delta}
&\qquad&{\cal O}_2 =  R^{\mu\nu\rho\sigma}R_{\mu\nu\rho}^{\;\;\;\;\;\;\delta} R^{\alpha\beta\gamma}_{\;\;\;\;\;\;\sigma}R_{\alpha\beta\gamma\delta}\\
&{\cal O}_3 = R^{\mu\nu\rho\sigma}R_{\mu\nu}^{\;\;\;\;\alpha\beta}R_{\alpha\beta}^{\;\;\;\;\gamma\delta}R_{\rho\sigma\gamma\delta} 
&\qquad&{\cal O}_4 = R^{\mu\nu\rho\sigma}R_{\mu\nu}^{\;\;\;\;\alpha\beta}R_{\rho\alpha}^{\;\;\;\;\gamma\delta}R_{\sigma\beta\gamma\delta}\\
&{\cal O}_5= R^{\mu\nu\rho\sigma}R_{\mu\nu}^{\;\;\;\;\alpha\beta}R_{\rho\;\;\alpha}^{\;\;\gamma\;\;\delta}R_{\sigma\gamma\beta\delta} 
&\qquad&{\cal O}_6 =  R^{\mu\nu\rho\sigma}R_{\mu\;\;\rho}^{\;\;\alpha\;\;\beta}R_{\alpha\;\;\beta}^{\;\;\gamma\;\;\delta}R_{\nu\gamma\sigma\delta}\\
&{\cal O}_7= R^{\mu\nu\rho\sigma}R_{\mu\;\;\rho}^{\;\;\alpha\;\;\beta}R_{\alpha\;\;\nu}^{\;\;\gamma\;\;\delta}R_{\beta\gamma\sigma\delta} .  &
\label{eq:EFT}
 \end{aligned}
 \ee
Note that linear dependences arise among operators as the dimension $D$ of spacetime decreases.  At quadratic order, $\mathcal{L}_2$ is unphysical in $D\leq 3$, a total derivative in $D=4$, and dynamical in $D\geq 5$.  Meanwhile, at quartic order, the number of algebraically independent operators $\mathcal{O}_i$ in $D=4,5,6,7,8$ is $2,4,6,6,7$, respectively \cite{Invariants}, with one linear combination---the eight-dimensional Euler density---a total derivative in $D = 8$ and hence dynamical only in $D\geq 9$ \cite{DeserSeminara}.

Our analysis hinges on the on-shell four-point graviton scattering amplitude, $M$, whose forward limit is intimately linked to the total cross-section by well-known analyticity arguments \cite{Pham:1985cr,IRUV}.   By marginalizing over the external graviton polarizations, we can then systematically derive a rigorous and inclusive set positivity bounds on the coefficients of operators in the graviton effective action.  Throughout, we assume a perturbative ultraviolet completion of gravity, so there exists a well-defined expansion in $\hbar$.

We begin with an analysis of quartic curvature corrections, proving that in $D=4$ the coefficients of the $(R^2)^2$ and $(R\tilde R)^2$ operators are positive.  Our results precisely match those of \Ref{Kleban}, which derived bounds from the condition of locally subluminal graviton propagation. We then generalize our arguments to $D=5$ and $D\geq 6$.  Subsequently, we obtain positivity constraints on supersymmetric operators in general $D$, which in the literature are sometimes denoted $t_8 t_8 R^4$ and $t_8 (R^2)^2$.  As a consistency check, we verify that all our constraints are satisfied in the bosonic, type II, and heterotic string theories.  

Moving on to quadratic curvature corrections, we argue that unitarity and analyticity exclude theories in which $\lambda \gg 1$ with no new degrees of freedom present at or below the mass scale $\Lambda \sim |\lambdaGB \kappa^2|^{-1/2}$, the natural cutoff associated with the Gauss-Bonnet term and the derivative expansion.  
Our results precisely accord with those of Maldacena et al. \cite{Maldacena}, who demonstrated that this class of theories is inconsistent with global causality.

The remainder of this paper is organized as follows. In \Sec{sec:Analyticity}, we review the arguments of \Ref{IRUV} whereby unitarity and analyticity impose rigorous positivity bounds on operator coefficients in effective theories.  Afterwards, in \Sec{sec:quartic} we apply these methods to establish the positivity of certain coefficients of quartic curvature operators, starting in $D=4$ and $D=5$ and then generalizing to $D\geq 6$.  We then apply our bounds to supersymmetric theories and string theories.  Finally, we study quadratic curvature operators in \Sec{sec:quadratic} and conclude in \Sec{sec:conclusions}.

\section{Analyticity Argument}\label{sec:Analyticity}

In this section we review how operator coefficients in effective field theories are constrained by the analyticity of scattering amplitudes at complex momenta.   
Our analysis follows that of \Ref{IRUV}, which derived positivity bounds on operator coefficients by relating the low-energy limit of forward amplitudes to strictly positive integrals of cross-sections.  

Our object of interest is the on-shell amplitude $M$ describing four-point graviton scattering in $D$ dimensions.  
Here,  the choice of the external polarizations is built into the functional form of $M$, as is the case for helicity amplitudes in $D=4$.
From this viewpoint, helicity is just a quantum number labeling the external states, no different from baryon or lepton number.  Sometimes it will be useful to view $M$ as a function of the external particle labels, $M=M(1,2,3,4)$, and other times as a function of 
the kinematic invariants, $M= M(s,t,u)$, where
\be  
s = -(k_1 + k_2)^2, \quad t = -(k_1 + k_3)^2, \quad u = -(k_1 + k_4)^2,
\ee
working in the convention where all momenta are incoming, so $k_1+k_2+k_3+k_4=0$. 

To derive constraints from analyticity, we will be interested in scattering amplitudes that are simultaneously forward and invariant under crossing in the $t$-channel.  Formally, $t$-channel crossing symmetry implies invariance under swapping
particle labels $1\leftrightarrow 3$ or $2\leftrightarrow4$ while leaving the functional form of $M$---which encodes the polarization choices---fixed, so
\be 
 \Msym(1,2,3,4) = \Msym(3,2,1,4) = \Msym(1,4,3,2) = \Msym(3,4,1,2). 
 \ee
 For external gravitons, 
  crossing symmetry is ensured if the exchanged states are indistinguishable.  This happens in $D=4$ if the states have identical helicity and more generally in $D$ dimensions if the states have the same polarization relative to their momenta. Mathematically, $M$ is crossing symmetric provided the momentum and polarization of particle 1 are related by an improper Lorentz transformation to those of particle 3, and likewise for particles 2 and 4.
In terms of kinematic invariants, a crossing symmetric amplitude then satisfies
\be 
M(s,t) = M(-s-t,t),
\label{eq:crossing relation}
\ee
where the momenta are swapped but the polarizations relative to momenta---which in $D=4$ are the helicities---are untouched.

Meanwhile, the forward limit of the amplitude, $M(s,t \rightarrow 0)$, corresponds to an identification of particles $1 \leftrightarrow 3$ and $2 \leftrightarrow 4$.  This is achieved simultaneously with crossing symmetry if we restrict to following kinematic regime:
\bea
\begin{array}{c}
\textrm{forward and} \\
\textrm{crossing symmetric} \\ 
\end{array} \implies \quad
 (k_3,\epsilon_3) \leftrightarrow (-k_1, \epsilon_1) &\;\;\textrm{and}\;\;& (k_4,\epsilon_4) \leftrightarrow (-k_2, \epsilon_2),
 \label{eq:cross forward}
\eea
where $\epsilon_1$ and $\epsilon_2$ are real linear polarizations. We choose a basis of linear polarizations because an amplitude with fixed external circular polarizations cannot be simultaneously crossing symmetric and forward.

   The forward and crossing symmetric amplitude, $M(s,t\rightarrow 0)$, can then be expanded in a power series in $s$ and $t$.  While analytic singularities in $s$ or $t$ arise, their form is severely restricted by the locality of the underlying theory.  As noted earlier, we assume throughout a perturbative ultraviolet completion of gravity, so we are justified in considering only the leading contribution in the $\hbar$ expansion, {\it i.e.}, tree-level exchange.
   
At tree level, analytic singularities in kinematic invariants enter at worst as simple poles.  Moreover, a $t$-channel singularity in the forward limit can only arise from non-local terms corresponding to graviton exchange induced by the leading Einstein-Hilbert interactions, so the general form for the forward amplitude is \be 
\Mforw(s,t\rightarrow 0) = \sum_{n=0}^\infty f_n s^n  + {\cal O}(s^2/t).
\label{eq:expand forward}
\ee
The first term is regular, as it is generated by heavy particle exchange, while the second term is singular because it comes from $t$-channel graviton exchange scaling as $\sim s^2 / t$.  The form of \Eq{eq:expand forward} together with the crossing symmetry relation of \Eq{eq:crossing relation} implies that
\be 
M(s,t\rightarrow 0) = 
M(-s,t\rightarrow 0)   + {\cal O}(s),
\label{eq:forward limit crossing}
\ee
where the first term arises because the limit of a regular function is the function evaluated at the limit of its arguments, while the second term is a residual contribution from applying crossing symmetry to the singular ${\cal O}(s^2/t)$ contribution.

The parameters $f_n$ depend on the coefficients of operators in the effective action of the low-energy theory.
To determine analyticity constraints, we consider the order $n$ residue of $M(s,t\rightarrow 0)$ in the complex $s$ plane, yielding
\be 
f_n = \frac{1}{2\pi i} \oint_\mathcal{C} \frac{\mathrm{d}s}{s^{n+1}} \left[ \Msym(s,t\rightarrow 0)  + {\cal O}(s^2/t)\right],\label{eq:littlecontour}
\ee
where $\mathcal{C}$ denotes a small contour encircling the origin.

As previously noted \cite{IRUV,IRWGC}, the ${\cal O}(s^2/t)$ singular contribution is formally infinite in the strictly forward limit and therefore a major obstacle to deriving bounds from analyticity.  Nevertheless, for $n\neq 2$ this term is eliminated by the residue theorem.  While forward singularities of order $s^n /t$ can arise from loop-level graviton exchange diagrams, we are working at leading order in the $\hbar$ expansion so these contributions are formally subdominant.  On the other hand,  $n=2$ is more subtle, but we will show that in certain parameter regimes the ${\cal O}(s^2/t)$ term can be subdominant to the rest of the amplitude, allowing for a bound to be placed.  In any case, we leave a detailed discussion of these issues for later sections and for now simply drop the ${\cal O}(s^2/t)$ contribution. Terms subleading in the forward limit of the Einstein-Hilbert amplitude must by power counting go as ${\cal O}(s)$ and will thus be eliminated in the contour integral for all $n\geq 2$.

By Cauchy's theorem, we can blow up $\mathcal{C}$ into a new contour $\mathcal{C}^\prime$ that runs just above and below the real $s$ axis, plus a circular boundary contour at infinity, yielding
\be
 f_n = \frac{1}{2\pi i} \left( \int_{-\infty}^{-s_0} + \int_{s_0}^\infty\right) \frac{\mathrm{d}s}{s^{n+1}} \Disc{\Msym (s,t\rightarrow 0)} + \text{boundary integral},
 \label{eq:total fn}
\ee
where $s_0$ is any scale above zero and below the first massive threshold in the ultraviolet completion. We note that $\Disc{\Msym (s,t\rightarrow 0)} = \Msym(s+i\epsilon,t\rightarrow 0) - \Msym(s-i\epsilon,t\rightarrow 0)$. By the Schwarz reflection principle, $\Msym(s^*,t\rightarrow 0) = \Msym(s,t\rightarrow 0)^*$, so
\be  
\Disc{\Msym (s,t\rightarrow 0)} = 2i \mathrm{Im}[\Msym(s,t\rightarrow 0)].
\label{eq:disc to im}
\ee  
The crossing symmetry relation in \Eq{eq:forward limit crossing} then implies that
\be 
\Disc{\Msym (-s,t\rightarrow 0)} = -\Disc{\Msym (s,t\rightarrow 0) },
\label{eq:disc forward}
\ee
dropping the ${\cal O}(s)$ term that arose from the ${\cal O}(s^2/t)$ singularity.

Throughout this paper we assume that $|\Msym(s)|$ is less divergent than $|s|^n$ at large $s$ so that the boundary term in \Eq{eq:total fn} can be dropped.\footnote{Strictly speaking, positivity bounds only require that the boundary term be non-negative, which is sometimes true given specific assumptions about the ultraviolet \cite{Low:2009di}.  We do not consider this possibility here. }  This is a physically reasonable assumption applicable to any ultraviolet completion in which the large $s$ behavior of the amplitude at fixed finite physical $t\ll s $ grows more slowly in $s$ than the Einstein-Hilbert contribution, which scales as $s^2/t$. A theory that fails this criterion would actually have worse ultraviolet behavior than Einstein gravity.  Operationally, this translates into the assumption that $|\Msym(s)|$ grows more slowly than $|s|^2$ at large $s$.  For example, this can be verified explicitly in the Regge behavior of string theory amplitudes, which scale as $s^{\alpha(t)}/t$ where $\alpha(t)<2$ for $t<0$ \cite{IRUV}.  

 Combining \Eq{eq:total fn} with \Eqs{eq:disc to im}{eq:disc forward} yields
\be 
\label{eq:dispersionrelation}
f_n = \frac{(-1)^n+1}{\pi }\int_{s_0}^\infty \frac{\mathrm{d}s}{s^{n+1}} \mathrm{Im}[{\Msym (s,t\rightarrow 0)}] .
\ee
For $n$ odd, this result trivially implies $f_n=0$ as required by crossing symmetry of $M$, while for $n$ even, it imposes a positivity condition.
In particular, we use the optical theorem to write $\mathrm{Im}[\Msym(s,t\rightarrow 0)] = s \sigma(s)$, where $\sigma(s)$ is the total cross-section.\footnote{While the total cross-section diverges in the presence of a $t$-channel singularity, $\textrm{Im}\,\Msym(s,t\rightarrow 0)$ and by extension $\sigma(s) = \textrm{Im}\,\Msym(s,t\rightarrow 0) /s$ are really proxies for the finite sum over all residues of heavy states in the complex $s$ plane.  By factorization, each contribution is positive since $M(hh \rightarrow hh)\sim - M(hh\rightarrow X)M(X\rightarrow hh)/(s-m^2+i\epsilon)$ on the $s$-channel resonance  of a massive state $X$ of mass $m$. }
 Crucially, in an interacting theory with new heavy states,  $\sigma(s)$ is strictly positive, so
\be 
f_n =  \frac{2}{\pi}\int_{s_0}^\infty \mathrm{d}s \frac{\sigma(s)}{s^{n}} > 0,\label{eq:coeffbound}
\ee
thus establishing positivity of $f_n$ for even $n$.

Here $f_n$ corresponds to the $s^n$ contribution to the low-energy amplitude, which is proportional to the operator coefficients of ${\cal L}_n$.  By power counting, we know that the low-energy amplitude can be expanded in powers of Mandelstam variables, so
\be 
M = \sum_{n=1}^\infty M_n ,
\ee
where the leading contribution arises from the Einstein-Hilbert action, which in the forward limit gives an amplitude
\be 
M_1(s,t\rightarrow 0) = -\epsilon_{1\mu\nu}\epsilon_1^{\mu\nu}
\epsilon_{2\rho\sigma}\epsilon_2^{\rho\sigma}\frac{\kappa^2 s^2}{t} + {\cal O}(s),
\label{eq:M1}
\ee
where the ${\cal O}(s)$ terms are regular in the forward limit.  
The remaining contributions $M_n$ are generated by ${\cal L}_n$.  In the subsequent sections, we derive precise analyticity bounds for the quartic and quadratic curvature corrections, ${\cal L}_4$ and ${\cal L}_2$.

\section{Bounds on Quartic Curvature Corrections}\label{sec:quartic}

In this section we derive bounds on ${\cal L}_4$, which encodes quartic curvature corrections to Einstein gravity.  The leading contributions from ${\cal L}_4$ are quartic graviton vertices, which contribute to graviton scattering amplitudes via contact interactions.  Since these corrections are free from kinematic singularities, their forward limit is regular.  Thus, to obtain a forward, crossing symmetric amplitude, we simply set $t= 0$, $\epsilon_3=\epsilon_1 $, and $\epsilon_4 =\epsilon_2$, as derived in \Eq{eq:cross forward}.   

After a lengthy calculation, we compute the quartic corrections to the graviton scattering amplitude in the forward limit to be
\be 
\begin{aligned}
M_4(s,t\rightarrow 0) &= \frac{\kappa^4 s^4}{2}   \big[(2 c_6 + c_7)(\epsilon_{1\mu\nu}\epsilon_1^{\;\;\mu\nu}\epsilon_{2\rho\sigma}\epsilon_2^{\;\;\rho\sigma}) + (32c_1 + 4c_2 + 2c_6) (\epsilon_{1\mu\nu} \epsilon_2^{\;\;\mu\nu})^2 \\&\hspace{-6mm} + (4c_2 + 16c_3 + 2c_6)(\epsilon_1^{\;\;\mu\nu}\epsilon_{2\nu\rho}\epsilon_1^{\;\;\rho\sigma}\epsilon_{2\sigma\mu}) + (4c_2 + 8c_4 + 2c_7)(\epsilon_1^{\;\;\mu\nu}\epsilon_{1\nu\rho}\epsilon_2^{\;\;\rho\sigma}\epsilon_{2\sigma\mu})\big].
\end{aligned}\label{eq:R4amp}
\ee
\Eq{eq:coeffbound} bounds $f_4$, corresponding to the coefficient of the $s^4$ contribution to the amplitude, to be positive.  To determine the constraint on the coefficients of ${\cal L}_4$, we should marginalize over all possible values of the independent polarizations, $\epsilon_1$ and $\epsilon_2$.  

To determine the full set of bounds, it will be convenient to map the question of positivity to a linear algebra problem.  To do so, we work in the center-of-mass frame, where the polarization tensors, $\epsilon_{1\mu}^{\;\;\;\;\nu}$ and $\epsilon_{2\mu}^{\;\;\;\;\nu}$ are real, symmetric $(D-2)$-by-$(D-2)$ matrices satisfying the usual tracelessness and normalization conditions,
\bea
\Tr( \epsilon_1) = \Tr( \epsilon_2) =0  \quad &\textrm{and}&\quad 
\Tr( \epsilon_1 \cdot \epsilon_1) = \Tr( \epsilon_2 \cdot \epsilon_2) =1.
\eea
Furthermore, we can define Hermitian matrices $H_+ = \{\epsilon_1,\epsilon_2\}/2$ and $H_-= i [\epsilon_1,\epsilon_2]/2$ encoding the polarization information, which enter the amplitude in terms of the invariants
\be 
x=\Tr(H_+)\Tr(H_+), \qquad y = \Tr(H_+\cdot H_+),\qquad z= \Tr(H_- \cdot H_-).
\label{eq:xyzdef}
\ee
We can then express the analyticity bound as
\be
(2 c_6 + c_7) + (32 c_1 + 4c_2 + 2c_6)x + (8c_2 + 16c_3 + 8c_4 + 2c_6 + 2c_7)y + (-16c_3 + 8c_4 -2c_6 + 2c_7)z > 0,\label{eq:xyzbound}
\ee
for all $(x,y,z)$ spanned by the graviton polarizations $\epsilon_1$ and $\epsilon_2$.  
What is the allowed space of $(x,y,z)$? An obvious set of necessary conditions are
\bea
0\leq x,y,z \leq 1 \quad &\textrm{and}&  \quad y+z \leq 1,
\eea
from familiar linear algebra inequalities.  In general $D$, finding the space spanned by the allowed $(x,y,z)$ is a highly nontrivial problem in matrix inequalities.   

In the next subsections, we will study various physically well-motivated scenarios, including general theories in $D=4$ and supersymmetric theories in arbitrary $D$.  

\subsection{Theories in $D=4$}

The number of linearly independent curvature invariants monotonically increases with the dimension of spacetime.  In $D=4$, there are only two independent quartic curvature invariants. Hence,  ${\cal L}_4$ in \Eq{eq:EFT} collapses to 
\be
{\cal L}_4
=  c_1 {\cal O}_1 + \tilde c_1 \tilde{\cal O}_1,\label{eq:L4D4}
\ee
where ${\cal O}_1$ is defined as in \Eq{eq:EFT} but $\tilde {\cal O}_1$ is unique to $D=4$, 
\bea
{\cal O}_1 =R^{\mu\nu\rho\sigma} R_{\mu\nu\rho\sigma}R^{\alpha\beta\gamma\delta}R_{\alpha\beta\gamma\delta} \quad &\textrm{and}&\quad \tilde{\cal O}_1 =R^{\mu\nu\rho\sigma}\tilde R_{\mu\nu\rho\sigma}R^{\alpha\beta\gamma\delta}\tilde R_{\alpha\beta\gamma\delta},
\label{eq:EFT4}
\eea
where $\tilde R_{\mu\nu\rho\sigma} = R_{\mu \nu}^{\;\;\;\;\alpha\beta} \epsilon_{\alpha\beta \rho\sigma}/2$ is the dual Riemann tensor.  The operator $\tilde {\cal O}_1$ can be written as a linear combination of any two the operators in \Eq{eq:EFT} modulo contributions proportional to $R$ and $R_{\mu\nu}$, which can be eliminated by the equations of motion.  For example,
\be 
\tilde{\cal O}_1 = 4 {\cal O}_2-4 {\cal O}_3 = - 4 {\cal O}_2+8{\cal O}_4 = \cdots, \label{eq:Otilde}
\ee
corresponding to a choice of operator coefficients, $(c_1, 4 \tilde c_1 , -4 \tilde c_1 ,0 ,0,0,0)$, $(c_1, -4 \tilde c_1 , 0 ,8 \tilde c_1  ,0,0,0)$, etc.  The ellipses in \Eq{eq:Otilde} denote equivalent representations in terms of other operators, which are not unique due to the linear dependence in $D=4$ of all but two of the operators in \Eq{eq:EFT}.  

In $D=4$, the invariants $(x,y,z)$ are constructed from real, symmetric, traceless 2-by-2 matrices, which we can parameterize by
\be 
\begin{aligned}
\epsilon_1 &= \vec \epsilon_1 \cdot \vec\sigma /\sqrt{2}\\
\epsilon_2 &= \vec \epsilon_2 \cdot \vec\sigma /\sqrt{2},
\end{aligned}
\ee 
where $\vec\epsilon_1$ and $\vec\epsilon_2$ are real unit polarization vectors and $\vec\sigma$ are the Pauli matrices.  Since $\epsilon_1$ and $\epsilon_2$ are real and symmetric, they only have components in $\sigma_1$ and $\sigma_3$, since $\sigma_2$ is imaginary and anti-symmetric.
From standard matrix identities, we see that 
$\{ \epsilon_1 , \epsilon_2 \} = \vec\epsilon_1 \cdot \vec\epsilon_2 $ and 
$[ \epsilon_1 , \epsilon_2 ] = i (\vec\epsilon_1 \times \vec\epsilon_2)\cdot \vec\sigma$.  Defining $\theta$  to be the angle between $\vec\epsilon_1$ and $\vec\epsilon_2$, we obtain
\be
(x,y,z) =   \cos^2 \theta \; (1,\tfrac{1}{2} ,0) + \sin^2\theta \; (0,0,\tfrac{1}{2} ) ,\label{eq:taus} 
\ee
 which defines an interval  whose endpoints are $(1,\frac{1}{2} ,0) $ and $(0,0,\frac{1}{2} )$. 
Inserting these $(x,y,z)$ values, along with the coefficient choice given by \Eqs{eq:L4D4}{eq:Otilde}, the bound \eqref{eq:xyzbound} takes the suggestive form
\be 
  c_1 \cos^2\theta + \tilde c_1 \sin^2\theta > 0,
\ee 
which obviously implies positivity of both coefficients separately,
\be 
c_1>0 \quad \textrm{and} \quad  \tilde c_1 >0,\label{eq:D4bounds}
\ee 
which correspond to parallel or perpendicular polarization vectors, respectively.
Our results exactly coincide with those derived from requiring subluminal graviton propagation \cite{Kleban}.

\subsection{Theories in $D=5$}

In $D=5$, there are four linearly independent quartic curvature invariants.  For the sake of generality we use the basis of \Eq{eq:EFT} with the linear dependences among operators assumed.  
For this analysis, we ascertain the physically allowed region for the invariants $(x,y,z)$, which in $D=5$ are constructed from real, symmetric, traceless 3-by-3 matrices. This requirement constrains $(x,y,z)$ to lie in the plane $1+2x-6y-2z=0$. Specifically,  $(x,y,z)$ are restricted to a planar triangular region,
\be
(x,y,z) = \sum_{i=1}^3 \tau_i v_i,\label{eq:5Dhull}
\ee
defined by three vectors
\be 
v_1 = (0,0,\tfrac{1}{2}), \qquad
v_2 = (1,\tfrac{1}{2},0), \qquad \text{and} \qquad
v_3 = (0,\tfrac{1}{6},0) \label{eq:5Dvertices}
\ee
for the real parameters $\tau_1,\tau_2,\tau_3\geq 0$ such that $\tau_1 +\tau_2 +\tau_3  = 1$.  The vertices \eqref{eq:5Dvertices} of this triangle can be reached by choices of physical polarizations.  In particular,
\be  
\begin{aligned}
v_1:\;\;\; &\epsilon_1 = \frac{1}{\sqrt{2}}\begin{pmatrix} 1 & 0 & 0 \\0 & -1 & 0 \\0 & 0 & 0 \end{pmatrix},&\;\;\;&\epsilon_2 = \frac{1}{\sqrt{2}}\begin{pmatrix} 0 & 1 & 0\\1 & 0 & 0 \\0 & 0 & 0 \end{pmatrix},\\
v_2:\;\;\; &\epsilon_1 = \frac{1}{\sqrt{2}}\begin{pmatrix} 1 & 0 & 0\\0 & -1 & 0\\0 & 0 & 0 \end{pmatrix},&\;\;\;&\epsilon_2 = \frac{1}{\sqrt{2}}\begin{pmatrix} 1 & 0 & 0\\0 & -1 & 0\\0 & 0 & 0 \end{pmatrix},\\
v_3:\;\;\; &\epsilon_1 = \frac{1}{\sqrt{2}}\begin{pmatrix} 1 & 0 & 0\\0 & -1 & 0\\0 & 0 & 0 \end{pmatrix},&\;\;\;&\epsilon_2 = \frac{1}{\sqrt{6}}\begin{pmatrix} 1 & 0 & 0\\0 & 1 & 0\\0 & 0 & -2 \end{pmatrix}.
\end{aligned}
\ee 
Plugging in \Eqs{eq:5Dhull}{eq:5Dvertices} back into \Eq{eq:xyzbound}, we obtain
\be 
\left[\begin{array}{rl}
&\tau_1(-8 c_3 + 4 c_4 + c_6 + 2 c_7)\\[.5em]
+&\tau_2(
 32 c_1 + 8 c_2 + 8 c_3 + 4 c_4 + 5 c_6 + 2 c_7) \\[.5em]
+&  \frac{1}{3}\tau_3(4 c_2 + 8 c_3 + 4 c_4 + 7 c_6 + 4 c_7) 
\end{array}\right] > 0,
\ee 
where we have repackaged the terms independent of $(x,y,z)$ in \Eq{eq:xyzbound} into the coefficients $\tau_{1},\tau_{2},\tau_{3}$ by re-expressing  $1$ as $\tau_1 +\tau_2 +\tau_3$. Thus, the necessary and sufficient set of bounds on quartic curvature corrections in $D=5$ are
\be 
\begin{aligned}
-8 c_3 + 4 c_4 + c_6 + 2 c_7&> 0   \\
32 c_1 + 8 c_2 + 8 c_3 + 4 c_4 + 5 c_6 + 2 c_7 &> 0  \\
4 c_2 + 8 c_3 + 4 c_4 + 7 c_6 + 4 c_7 &> 0,
\end{aligned}\label{eq:D5bounds}
\ee 
coming from analyticity of the four-point graviton scattering amplitude.

\subsection{Theories in $D\geq 6$} 
 
Consider finally the general case of $D\geq 6$.  It is a non-trivial linear algebra problem to determine the parameter space of $(x,y,z)$ corresponding to physical polarization configurations. Each physically allowed point $(x,y,z)$ yields a positivity bound via \Eq{eq:xyzbound}.  The set of all positive linear combinations of such bounds is given by plugging into \Eq{eq:xyzbound} the set of all points in the convex hull $S$ spanning physically allowed values of $(x,y,z)$. Fully characterizing all such $(x,y,z)$ is beyond the scope of the present work.  However, we can derive a general collection of necessary conditions from a subset of extremal vertices on the boundary of $S$.
The details of the calculation are given in \App{Matrices}, but the vertices are
\be  
 \begin{aligned}
v_1 &= \left(0,0,\tfrac{1}{2}\right)\\
v_2 &= \left(1,1-\tfrac{3}{D-2} + \tfrac{1}{D-3},0\right)\\
v_3 &= \left(0,\tfrac{D-4}{2(D-2)},0\right)\\
v_4 &= \left(1,\tfrac{1}{D-2} \left[1+ \tfrac{4(D\!\!\!\!\mod 2)}{(D-1)(D-3)} \right],0\right)\\
v_5 &= \left(0,0,0\right).
\end{aligned}\label{eq:vertices}
 \ee
These vectors can be realized by physical polarization choices. The bounds associated with the $(x,y,z)$ values in \Eq{eq:vertices} are necessary for analyticity of four-point scattering amplitudes and moreover are a subset of the minimal basis of sufficient bounds. Numerical evaluation, via the explicit computation of $x$, $y$, and $z$ for pseudorandom,
traceless, unit-norm matrix pairs of various dimensions, shows that the convex hull defined by the vertices in \Eq{eq:vertices} is in fact equal to the full hull $S$ for even $D$ but is slightly smaller than $S$ in odd $D$. 
Note that the vectors~\eqref{eq:vertices} are a generalization of those we saw in earlier sections, so $v_1$, $v_2$, and $v_3$ coincide with the vectors from $D=5$.  Moreover, each corner corresponds to a certain extreme configuration of polarizations.  
For example, $v_1$ corresponds anticommuting polarizations as in \Eq{eq:5Dvertices}, while $v_2$, $v_3$, $v_4$, and $v_5$ correspond to commuting polarizations.    For the latter, the polarizations are mutually diagonalizable and can without loss of generality be represented as traceless diagonal matrices. See \App{Matrices} for details.

Plugging the vectors in \Eq{eq:vertices} into the bound in \Eq{eq:EFT}, we obtain the positivity bounds 
\be 
 \begin{aligned} 
-8c_3 + 4c_4 + c_6 + 2c_7 &> 0\\
2\left(1 - \tfrac{3}{D-2} + \tfrac{1}{D-3}\right) (4c_2 + 8c_3 + 4c_4 +c_6 + c_7)
+32c_1 + 4c_2 + 4c_6 + c_7 &> 0 \\
\left(\tfrac{D-4}{D-2}\right)(4c_2 + 8c_3 + 4c_4 +c_6 + c_7) + 2c_6 + c_7&> 0 \\
\left(\tfrac{2}{D-2}\right)\left[1 + \tfrac{4(D\!\!\!\!\mod 2)}{(D-1)(D-3)}\right] (4c_2 + 8c_3 + 4c_4 +c_6 + c_7)+32c_1 + 4c_2 + 4c_6 + c_7 &> 0 \\
2c_6 + c_7&> 0,
\end{aligned}\label{eq:D6bounds}
\ee
which are a stringent set of requirements on quartic curvature corrections to general relativity in $D\geq 6$, necessary to guarantee analyticity of scattering amplitudes. 

\subsection{Supersymmetric Theories}

We now consider supersymmetric quartic curvature corrections.  Conveniently, \Ref{Roo} derived a basis for independent off-shell supersymmetric quartic curvature invariants, 
\be 
{\cal L}_4
= A {\cal O}_A +  B {\cal O}_B+ C {\cal O}_C,
\ee
where ${\cal O}_A$, ${\cal O}_B$, and ${\cal O}_C$ are proportional to more familiar looking forms denoted in the literature \cite{Tseytlin,Roo} by $t_8 t_8R^4$, $t_8 (R^2)^2$, and $\epsilon_{10} \epsilon_{10} R^4$, respectively.  In terms of the basis defined in \Eq{eq:EFT}, these supersymmetric operators are
\be 
\begin{aligned}
{\cal O}_A &= {\cal O}_1 -16{\cal O}_2 + 2{\cal O}_3  -32 {\cal O}_5 + 16 {\cal O}_6 +32 {\cal O}_7\\
{\cal O}_B &= -{\cal O}_1 +  8 {\cal O}_2-2{\cal O}_3 +4{\cal O}_4 \\
{\cal O}_C &=  {\cal O}_1 -16{\cal O}_2 + 2{\cal O}_3 +16 {\cal O}_4 -32 {\cal O}_5 + 16 {\cal O}_6 -32 {\cal O}_7,
\end{aligned}
\ee
corresponding to the following choice of operator coefficients:
\be 
\begin{aligned}
c_1 &=A-B+C && \;\;\;\;\; \;\;& c_2 &= -16A+8B-16C &&\;\;\;\;\;\;\;& c_3 &=2A-2B+2C\\
c_4 &= 4B+16C &&& c_5 &= -32A-32C &&& c_6 &= 16A+16C\\
c_7 &= 32A-32C.
\end{aligned}
\ee
Plugging this choice into \Eq{eq:xyzbound}, we obtain
\be 
A + B(y+z) > 0.\label{eq:SUSYboundyz}
\ee 
Note that $C$ drops out of the calculation completely, since at quartic order in graviton perturbations it is a total derivative in all dimensions \cite{DeserSeminara}.  Recall from \App{Matrices} that while simple matrix identities imply that $y+z \leq 1$, no point in the hull $S$ actually saturates this bound. For example, in $D=4$, Eq.~\eqref{eq:taus} implies that $y+z =\frac{1}{2}$, so $A+\frac{1}{2}B>0$. In $D=5$, \Eq{eq:5Dvertices} implies that $\frac{1}{6} \leq y+z \leq \frac{1}{2}$, so $A+\frac{1}{6}B>0$ and $A+\frac{1}{2}B>0$. Finally, in $D\geq 6$, inputting the vertices in \Eq{eq:vertices} into \Eq{eq:SUSYboundyz} yields the complete set of positivity bounds for quartic curvature operators in supergravity theories.  In summary, we find
\be 
\begin{aligned}
A + \tfrac{1}{6}B &> 0 \qquad && (D=5) \\
A  &>  0 \qquad &&(D\geq 6) \\
A+\left(1-\tfrac{3}{D-2}+\tfrac{1}{D-3}\right)B  &>  0, \qquad &&({\rm any\;} D)
\label{eq:SUSYbound}
\end{aligned}
\ee
noting that in $D=4$ and $D=5$, the final bound reduces to $A+\frac{1}{2}B>0$.

\subsection{String Theories}

As a consistency check, we now apply our bounds to string theory, which is arguably the leading candidate for the ultraviolet completion of gravity.  Conveniently, quartic curvature corrections have been been dutifully computed at tree-level in the existing literature for the bosonic \cite{Jack,Jack2}, type II \cite{Gross+Witten,Gross+Sloan,Kikuchi}, and heterotic string \cite{Gross+Sloan,Kikuchi}. The type I string is dual to the heterotic string and has the same low-energy effective action \cite{BBS,Tseytlin}, so we need not consider it as a separate case. The resulting effective theory is described by
\be 
{\cal L}_4 = A {\cal O}_A +B {\cal O}_B +C {\cal O}_C + \Delta {\cal O}_{\Delta},
\ee
where ${\cal O}_A$, ${\cal O}_B$, and ${\cal O}_C$ are the supersymmetric operators from the previous section and ${\cal O}_{\Delta}$ is a non-supersymmetric operator defined as
\be 
{\cal O}_{\Delta} = - {\cal O}_1  +10 {\cal O}_2 +{\cal O}_4 .
\ee 
In various string theories, the operator coefficients are
\bea
\begin{tabular}{c|c|c|c|c|}
    & $A$ & $B$ & $C$ & $\Delta$ \\ \hline
bosonic &  $\zeta(3)$  & 0 & $-\zeta(3)$ & $16$   \\ \hline
  type II &  $\zeta(3)$  & 0 & $-\zeta(3)$ & 0 \\ \hline
heterotic & $\zeta(3)$   & $1$ & $-\zeta(3)$ & 0  \\ \hline
\end{tabular}\label{eq:stringcoeffs}
\eea
where each entry is normalized by a factor of $\alpha^{\prime 3}/1024\kappa^2$.

As expected, since the type II and heterotic string theories are supersymmetric, their coefficients in \Eq{eq:stringcoeffs} satisfy the bound for supersymmetric theories in \Eq{eq:SUSYbound}.  Since the bosonic string is non-supersymmetric, the bound is more complicated.  In particular, plugging the corresponding operator coefficients into \Eq{eq:xyzbound}, we obtain
\be 
\zeta(3) + 2(x+11y +z)  > 0,
\ee 
which is indeed positive, as $x,y,z \geq 0$.  Thus, we have verified that quartic curvature corrections in bosonic, type II, and heterotic string theory are consistent with unitarity and analyticity.

\section{Bounds on Quadratic Curvature Corrections} \label{sec:quadratic}

Next, we consider analyticity constraints on ${\cal L}_2$, which characterizes quadratic curvature corrections in the graviton effective theory.  
As shown in \Ref{Zwiebach}, the Gauss-Bonnet term
\be 
\mathcal{L}_2 = \lambdaGB (R_{\mu\nu\rho\sigma}R^{\mu\nu\rho\sigma}-4R_{\mu\nu}R^{\mu\nu} + R^2)\label{eq:GB} 
\ee  
does not introduce ghost modes in any dimension $D$, so in this basis the graviton propagator is unmodified.  To avoid ghost pathologies, we only consider curvature invariants of this form.  For $D=4$, the Gauss-Bonnet term is furthermore a total derivative and thus does not affect local dynamics.  As recently shown \cite{evanescent}, however, the Gauss-Bonnet term is critical for computing and interpreting the leading ultraviolet divergences of pure gravity.

Expanding to leading order in the Gauss-Bonnet coefficient $ \lambdaGB$, we compute the quadratic curvature correction to the graviton scattering amplitude in the forward limit,
\be
\begin{aligned}
M_2(s,t\rightarrow 0) &= 4\lambdaGB \kappa^4   s^2  \Bigl[\epsilon_{1}^{\mu\nu}\epsilon_{3\mu\nu}\epsilon_{2}^{\rho\sigma}\epsilon_{4\rho\sigma}+\epsilon_{1}^{\mu\nu}\epsilon_{3\nu\rho}\epsilon_{2}^{\rho\sigma}\epsilon_{4\sigma\mu}+\epsilon_{1}^{\mu\nu}\epsilon_{3\nu\rho}\epsilon_{4}^{\rho\sigma}\epsilon_{2\sigma\mu}\\ &\qquad\qquad\qquad\qquad +
\frac{2}{t}\left(k_{2}^{\mu}k_{4}^{\nu}\epsilon_{2\nu}^{\;\;\;\rho}\epsilon_{4\rho\mu}\epsilon_{1\alpha\beta}\epsilon_{3}^{\alpha\beta}+k_{1}^{\mu}k_{3}^{\nu}\epsilon_{1\nu}^{\;\;\;\rho}\epsilon_{3\rho\mu}\epsilon_{2\alpha\beta}\epsilon_{4}^{\alpha\beta}\right)\Bigr],
\end{aligned}\label{eq:R2amp1}
\ee
where we have expanded formally in $t$-dependence arising from propagator denominators, but we have yet to evaluate the numerators in the forward limit.

The first line of \Eq{eq:R2amp1} is manifestly regular in the forward limit $t=0$, so for these terms we can simply set $\epsilon_3 = \epsilon_1$ and $\epsilon_4 = \epsilon_2$.
On the other hand, the second line of \Eq{eq:R2amp1} is naively singular since $1/t$ diverges as $t\rightarrow 0$.  However, this singularity is canceled by the numerator factor, which vanishes in the forward limit as $\epsilon_3 \rightarrow \epsilon_1$ and $\epsilon_4 \rightarrow \epsilon_2$.  It will be convenient to rewrite this expression in terms of the momentum transfer,
\be 
q=k_1 + k_3=-(k_2 + k_4),\label{eq:exchangedmomentum}
\ee
where $t=-q^2$. For real kinematics, $q$ is spacelike and vanishes in the forward limit. Note that $q^\mu q^\nu/q^2$ is simply a projection operator in the direction of the spacelike exchanged momentum. 

 We note that $k_3$ is simply a real spatial rotation of $-k_1$, and likewise for $k_4$ and $k_2$. By symmetry, this then implies that $\epsilon_1^{\mu\nu}k_{3\nu} = \epsilon_3^{\mu\nu}k_{1\mu} = \epsilon_1^{\mu\nu}q_\mu$ and $\epsilon_2^{\mu\nu}k_{4\mu}=\epsilon_4^{\mu\nu}k_{2\mu} = -\epsilon_2^{\mu\nu}q_\mu$ at leading order in $q$. Rewriting \Eq{eq:R2amp1} in terms of $q$, we then have
\be
\begin{aligned}
M_2(s,t\rightarrow0)&=  4\lambdaGB \kappa^4  s^2 \Bigl[\epsilon_{1\mu\nu}\epsilon_1^{\;\;\mu\nu}\epsilon_{2\rho\sigma}\epsilon_2^{\;\;\rho\sigma} +2 \epsilon_1^{\mu\nu}\epsilon_{1\nu\rho}\epsilon_2^{\rho\sigma}\epsilon_{2\sigma\mu} \\ &\qquad\qquad\qquad\qquad   -\frac{2q^\mu q^\nu}{q^2}\left( \epsilon_{2\mu}^{\;\;\;\;\rho}\epsilon_{2\rho\nu} \epsilon_{1\alpha\beta}\epsilon_1^{\;\;\alpha\beta}+\epsilon_{1\mu}^{\;\;\;\;\rho}\epsilon_{1\rho\nu}  \epsilon_{2\alpha\beta}\epsilon_2^{\;\;\alpha\beta}\right) \Bigr],\label{eq:R2amp2}
\end{aligned}
\ee
which is regular because the projection operator $q^\mu q^\nu /q^2$ is finite in the forward limit.
To obtain a bound on $ \lambdaGB$, we consider all possible choices for the external momenta and polarizations and impose positivity bounds on the forward amplitude in \Eq{eq:R2amp2}.

As expected, quadratic curvature corrections to graviton scattering scale as $M_2 \sim  \lambdaGB\kappa^4  s^2$, so to extract an analyticity bound we should apply \Eq{eq:coeffbound} for a second-order residue, corresponding to $n=2$.  Unfortunately, this choice also extracts the $t$-channel singular contribution from leading-order graviton exchange, $M_1 \sim -\kappa^2 s^2 / t$.  In the forward limit, this contribution is formally infinite. Of course, in any physical experiment there is an infrared scale $\mu$ that regulates these contributions from long distance physics.  This would arise, {\it e.g.}, from a finite detector resolution or beam width \cite{Melnikov}.   As is common practice for infrared divergences in scattering amplitudes, we introduce a mass regulator, sending $t\rightarrow t-\mu^2$ in the denominator.  This approach was also used in \Ref{IRUV} to make sense of a theory of interacting massless scalars with trilinear couplings.  Note that as in gauge theory, the mass $\mu^2$ is a formal regulator that leaves the number degrees of freedom untouched---so the vDVZ discontinuity \cite{vDV,Z}, which arises for a physical graviton mass included via a Fierz-Pauli Lagrangian term, does not apply here.  

While $\mu^2$ tames the formal infrared divergences, for $\lambdaGB \lesssim 1$ the forward amplitude will be dominated by finite but large contributions from Einstein-Hilbert interactions because $|M_1 |\gg | M_2|$ in this regime.\footnote{Note that taking $t$ strictly to zero is not required to derive a positivity bound \cite{Martin} and positivity  holds for any non-negative $t$ below $\mu^2$ \cite{Nicolis:2009qm}. However, we will not need this more general result for our purposes.}    However, by explicit calculation, we can see from \Eq{eq:M1} that $M_1 \sim +\kappa^2 s^2 / \mu^2$, which is positive.  So while positivity is satisfied, we learn nothing beyond what is already borne out from scattering via the leading Einstein-Hilbert term.  

To place a bound on the coefficient $\lambdaGB$, we must then restrict to a parameter regime where $|M_1 |\lesssim | M_2|$, so  the contributions from graviton exchange are subdominant to those from the Gauss-Bonnet term.   This implies that $1/\mu^2 \lesssim | \lambdaGB  \kappa^2|$.  Together with the requirement that $|s| \gg \mu^2$, necessary to treat $\mu$ as a regulator, this forces us to consider the regime
\be 
\sqrt{|s|} \gg \mu \gtrsim  \Lambda,
\label{eq:t bound}
\ee 
where $ \Lambda \sim | \lambdaGB \kappa^2|^{-1/2}$ is the scale of the would-be natural cutoff associated with the derivative expansion.  We assume throughout that $\Lambda \ll \kappa^{\frac{2}{2-D}}$ so that it is below the Planck scale in $D$ dimensions.

Of course, \Eq{eq:t bound} points to a naively pathological region of the effective field theory, given the reasonable expectation of new degrees of freedom of mass $m$ where $ m \sim \Lambda$. Moreover, \Eq{eq:t bound} indicates that the infrared regulator $\mu$ must be larger than some other energy scale $\Lambda$. Nevertheless, one can {\it a priori} envision an ultraviolet completion in which $m \gg \Lambda$, so new degrees of freedom enter at a parametrically higher scale. In that case, $\Lambda$ is not the scale of any physical states in the theory and is merely the combination of parameters that appears in the higher-dimension operator. Indeed, at the level of the scattering amplitude, there are no discontinuities that appear around $\Lambda$ to signal new degrees of freedom.

Thus, $\mu$ remains smaller than any physical mass scale in the theory and indeed can be treated consistently as an infrared regulator. In the absence of new states at $\Lambda$, the Gauss-Bonnet term acts effectively as a primordial contact operator over a wide range of scales.
Precisely such a scenario was considered in \Ref{Maldacena}, where it was found that such a low-energy effective theory is acausal without new states at or below $m\lesssim\Lambda$.  Other authors \cite{SarkarWall} have likewise argued that a pure Gauss-Bonnet theory is inconsistent with black hole thermodynamics.  We will likewise find a pathology in this theory coming from unitarity and analyticity.

To apply constraints from unitarity and analyticity, we must first ensure that the low-energy theory is sensible enough that we can even speak of a long-distance scattering amplitude.  Indeed, \Eq{eq:t bound} is plainly strange since $|s| \gg \Lambda^2$ violates the derivative expansion.    This was required in order for the Gauss-Bonnet interactions to dominate over the Einstein-Hilbert action, as was also assumed in \Ref{Maldacena}.  Naively, one would expect a gross departure from perturbative unitarity, {\it e.g.}, probability amplitudes greater than one as well as a breakdown of the loop expansion.  Nevertheless, there is a wide range of scales where neither sickness actually arises.  This hinges on the fact that the theory depends on $\Lambda$ as well as $\kappa$, the gravitational coupling constant.

In particular, note that amplitudes can still be perturbatively small in the regime specified by \Eq{eq:t bound}.  For example, $M_2 \sim \kappa^2 s^2 / \Lambda^2$ is still sensible provided $\kappa$ is sufficiently small, corresponding to the weak gravity limit.  We can make this more precise by considering the leading effect of the Gauss-Bonnet term, which is a cubic vertex of the schematic form $\lambda \kappa^3 \partial^4 h^3$.  Inserting this vertex into low-energy amplitudes, we find that the theory remains under perturbative control provided $\lambda \kappa^3$ ($\sim \kappa/\Lambda^2$) times the appropriate powers of energy is sufficiently small.   In $D$ dimensions this implies that 
\be 
|s| \ll \left( \frac{\Lambda^2}{\kappa} \right)^{\frac{4}{2+D}}
\label{eq:pertunit}
\ee
to safely reside within the regime of perturbativity.\footnote{A similar statement applies to pions, which have quartic vertices of the form $\partial^4 \pi^4 / \Lambda^2 v^2$ where $v$ is the breaking scale and $\Lambda$ controls the derivative expansion.  The theory maintains perturbative control provided $s \ll \Lambda v$. }  Moreover, \Eq{eq:pertunit} also ensures a perturbative loop expansion, since radiative corrections always introduce additional insertions of the Gauss-Bonnet interactions.

For our purposes, we assume a weak gravity limit defined by \Eq{eq:pertunit}, so the low-energy theory is perturbatively unitary.  When then apply the method of \Sec{sec:Analyticity}, where the contour around the origin in the complex $s$ plane is widened so as to satisfy \Eq{eq:t bound}, ensuring that $s$ is large compared to the scale of the infrared cutoff and that the Gauss-Bonnet term dominates the amplitude.   Note also that the initial contour encircles a region below the heavy particle threshold, $m \gg \Lambda$.

To see how a pathology arises, it will be convenient to define coordinates transverse to the incoming momenta, $(x_1, x_2, \ldots x_{D-2})$.  Without loss of generality, we take the forward limit such that the infinitesimal momentum transfer lies in the $x_1$ direction, which we henceforth refer to as the ``direction of approach." In turn,  $q^\mu q^\nu /q^2 $ is a projection operator onto this direction.  Next, we define a particular subset of polarizations in the transverse plane, defined by rank-two diagonal matrices of the form
\bea
d^{(i,j)} &=& \frac{1}{\sqrt{2}} \; {\rm diag}(0,\ldots, 0, \overset{x_i}{\overbrace{1}}, 0,\ldots,0, \overset{x_j}{\overbrace{-1}} ,0,\ldots),\label{eq:dij}
\eea
with zero entries except in the $x_i$ and $x_j$ directions.  As the only preferred direction is $x_1$, labeling the direction from which we approach the forward limit, the relevant physical polarizations are $d^{(1,2)}$, $d^{(2,3)}$, and $d^{(3,1)}$.  The forward limit of the quadratic correction to the graviton scattering amplitude in \Eq{eq:R2amp2} for various polarization combinations is
\be 
\begin{aligned}
M(s,t\rightarrow 0) &=& 
2\lambdaGB \kappa^4  s^2\times\left\{
\begin{array}{ll}
0 ,&\quad \epsilon_1 = \epsilon_2 = d^{(1,2)} \\[0.5em]
4 ,&\quad\epsilon_1 = \epsilon_2 = d^{(2,3)} \\[0.5em]
-1,&\quad  \epsilon_1  =d^{(1,2)} \textrm{ and } \epsilon_2  =d^{(1,3)}. \\
\end{array}
\right.
\label{eq:MGBbound}
\end{aligned}
\ee
In the first case, $\epsilon_1 = \epsilon_2 = d^{(1,2)}$, corresponding to polarizations that have support in the direction of approach.  In the case of $D=4$, this is required because the transverse space only has two dimensions.  As expected, the amplitude vanishes in this regime because the Gauss-Bonnet term is a total derivative in $D=4$.  Meanwhile, the second case, $\epsilon_1 = \epsilon_2 = d^{(2,3)}$ occurs when both polarizations are orthogonal to the direction of approach. Of course, this requires dimensions $D \geq 5$.    Finally, in the last case, $\epsilon_1  = d^{(1,2)}$ and $\epsilon_2  = d^{(1,3)}$, the polarizations occupy different planes but share support in the direction of approach, which is only possible in $D\geq 5$.

The upshot of \Eq{eq:MGBbound} is that in $D\geq 5$, different polarization configurations can yield opposite signs for the corrections to the forward scattering amplitude.  As a result, this excludes both signs of $\lambdaGB$ and thus forbids it entirely.  Of course, we made the assumption of \Ref{Maldacena} that the Gauss-Bonnet term is an effectively primordial contact operator insofar as new degrees of freedom enter only at a scale far above the naive cutoff.  Hence, the positivity violation in \Eq{eq:MGBbound} simply implies that this assumption is false.  We conclude that a primordial Gauss-Bonnet term is forbidden and new degrees of freedom are required at or below the cutoff $\Lambda$.

\section{Conclusions} \label{sec:conclusions}
In this paper, we have derived rigorous bounds on the coefficients of quartic and quadratic curvature corrections in the low-energy effective theory of gravitons. Our results hinge on very general principles: quantum mechanical unitarity and analyticity of scattering amplitudes.  Consequently, these constraints apply to any consistent perturbative ultraviolet completion of gravity. For the quartic curvature operators defined in Eqs.~\eqref{eq:L4}, \eqref{eq:EFT}, \eqref{eq:L4D4}, and \eqref{eq:EFT4}, we derived the positivity bounds in \Eq{eq:D4bounds} in $D=4$, \Eq{eq:D5bounds} in $D=5$, and \Eq{eq:D6bounds} in arbitrary $D\geq 6$. We also presented constraints on supergravity theories and checked that all of our results are consistent with known calculations in weakly-coupled string theories. 
For the quadratic curvature correction in \Eq{eq:GB}, we showed that both signs of its coefficient $\lambdaGB$ are inconsistent unless new degrees of freedom enter at the natural cutoff $\Lambda\sim|\lambda\kappa^2|^{-1/2}$ specified by the effective theory.  In short, a primordial Gauss-Bonnet term is forbidden.

Many possibilities remain for future work. While four-point graviton scattering cannot probe curvature operators beyond quartic order, little is known of higher-point amplitudes.  Such amplitudes are functions of many more kinematic invariants and should thus enforce commensurately more positivity constraints.  Another issue meriting further study is that of cubic curvature operators.  Here, positivity bounds encounter technical challenges due to the vanishing of the associated amplitude in the forward limit \cite{Nicolis:2009qm, Elvang}.  As noted in \Ref{Elvang}, this problem is closely related to the $a$-theorem in $D=6$. 

Distinguishing low-energy effective theories that are consistent with ultraviolet completion from those that are not presents a significant challenge. Systematizing this procedure is important for delineating the space of possible physical laws, but has also become important for model-building in more phenomenological contexts \cite{Naturalness} and in inflation \cite{Baumann:2015nta,Croon:2015fza,Reece,Raman,Shiu}. 
In this paper, the low-energy tools of analyticity and unitarity enabled us to find solutions to this problem in gravitational theories, allowing us to constrain higher-curvature corrections to gravity in our own universe---applying our quartic curvature results to $D=4$---and further discover bounds applicable in any consistent theory.

\begin{center} 
 {\bf Acknowledgments}
 \end{center}
 \noindent 
We thank Nima Arkani-Hamed and John Schwarz for useful discussions and comments. B.B. is supported in part by the MIUR-FIRB grant RBFR12H1MW, by the Agence Nationale de la Recherche under contract ANR 2010 BLANC 0413 01, and by the ERC Starting Grant agreement 278234 ``NewDark" project. C.C.~is supported by a Sloan Research Fellowship and a DOE Early Career Award under Grant No.~DE-SC0010255.  G.N.R.~is supported by a Hertz Graduate Fellowship and a NSF Graduate Research Fellowship under Grant No.~DGE-1144469.
B.B. and C.C. thank the Aspen Center for Physics and the Munich Institute for Astro- and Particle Physics (MIAPP) of the DFG cluster of excellence ``Origin and Structure of the Universe" for their hospitality and support while this work was in progress.

\appendix
\section{Bounding Invariants in General Dimension}\label{Matrices}
\setcounter{equation}{0}
\numberwithin{equation}{section}

We have shown that the graviton scattering amplitude can be expressed in terms of invariant products of graviton polarizations $\epsilon_1$ and $\epsilon_2$, which are real, symmetric, traceless $(D-2)$-by-$(D-2)$ matrices with unit normalization.  To recapitulate from \Sec{sec:quartic}, given the Hermitian matrices 
\bea
H_+ = \{\epsilon_1,\epsilon_2\}/2 \quad &\textrm{and} & \quad H_- = i[\epsilon_1,\epsilon_2]/2,
\eea
we can define the invariants
\be 
x=\Tr(H_+)\Tr(H_+), \qquad y = \Tr(H_+\cdot H_+),\qquad z= \Tr(H_- \cdot H_-).
\ee
The space of physical polarizations $\epsilon_1$ and $\epsilon_2$ then maps onto a physical region in $(x,y,z)$, which through \Eq{eq:xyzbound} implies positivity constraints on operator coefficients in the effective theory.   

What are the bounds on $(x,y,z)$?  We first note that since $H_+$ and $H_-$ are Hermitian, their squares are positive semidefinite, so $x,y,z\geq 0$. Moreover, a straightforward application of the Cauchy-Schwarz inequality implies $x = \Tr(\epsilon_1 \cdot \epsilon_2)\Tr(\epsilon_1 \cdot \epsilon_2) \leq \Tr(\epsilon_1 \cdot \epsilon_1)\Tr(\epsilon_2 \cdot \epsilon_2)= 1$, with equality if and only if $\epsilon_1 = \pm \epsilon_2$, and similarly $y+z = \Tr(\epsilon_1\cdot \epsilon_1 \cdot \epsilon_2\cdot \epsilon_2) \leq  \Tr(\epsilon_1 \cdot \epsilon_1)\Tr(\epsilon_2 \cdot \epsilon_2) = 1$.  There are, however, many additional constraints on $(x,y,z)$, which we now discuss.

Crucially, a weighted average of any number of positivity bounds yields another valid positivity bound.  This implies that a space of necessary conditions can be derived by constructing a convex hull $S$  in $(x,y,z)$ that contains the physically allowed region.  Without loss of generality, $S$ is 
\be 
S =  (x,y,z) = \left\lbrace \sum_{i=1}^n \tau_i v_i \;\Bigg\vert\; \tau_i \geq 0 \textrm{ and } \sum_{i=1}^n \tau_i = 1 \right\rbrace,
\ee 
where $v_i$ denote extremal points. In this Appendix, we will construct the subset of the $v_i$ that are on the edges of the unit cube in $(x,y,z)$; let the convex hull described by these vertices be $\tilde{S}$. In even dimension, numerical evaluation suggests that $S=\tilde{S}$, while in odd $D>6$, it is possible for points to lie slightly outside $\tilde{S}$. 

Let us first consider the case where $\epsilon_1$ and $\epsilon_2$ are anticommuting, so $x=y=0$ and we wish to maximize $z$. Going to a basis in which $\epsilon_1$ is diagonal, we find that antisymmetry of $\epsilon_1 \cdot \epsilon_2$ implies that for each $i,j$,
\be
(\epsilon_{1ii} + \epsilon_{1jj})\epsilon_{2ij} = 0 \label{eq:anticommuting}
\ee
and 
\be
z = -\Tr(\epsilon_1 \cdot \epsilon_2 \cdot \epsilon_1\cdot  \epsilon_2) = \sum_{i,j} \epsilon_{1ii}^2 \epsilon_{2ij}^2.\label{eq:zxy0}
\ee
Since \Eq{eq:anticommuting} implies $\epsilon_{1ii} \epsilon_{2ii} = 0$ for each $i$, the normalization condition $\sum_{i,j} \epsilon_{2ij}^2 = 1$ implies by \Eq{eq:zxy0} that nonzero diagonal terms in $\epsilon_2$ can only decrease $z$. We therefore take $\epsilon_2$ to have vanishing diagonal. Similarly, since $\sum_i \epsilon_{1ii}^2$ is fixed to unity, we should require that, for each $i$ for which $\epsilon_{1ii} \neq 0$, there exists $j$ such that $\epsilon_{2ij} \neq 0$; letting $\epsilon_{1i_0 i_0}$ be nonvanishing for some $i_0$ even if $\epsilon_{2i_0 j} = 0$ for all $j$ would decrease $z$ by \Eq{eq:zxy0}. Writing $\sum_{j} \epsilon_{2ij}^2 = \rho_i$, where $\sum_i \rho_i = 1$ by the normalization constraint, we can then consider $z$ to be a weighted average over the $\epsilon_{1ii}^2$.
Thus, $z$ is maximized when we weight the average most in favor of the $i$ for which $\epsilon_{1ii}^2$ is maximal. Suppose there are $N$ such $i$, which we can without loss of generality take to be 1 through $N$, for which $\epsilon_{1ii}^2$ takes its maximal value, {\it i.e.}, $\epsilon_{1i^* i^*}^2 = \max_i \epsilon_{1ii}^2 \equiv \varepsilon^2$ for all $i^* \in \{1,...,N\}$. Then $z$ is maximized when we have $\rho_i = 1/N$ for $i\in \{1,...,N\}$ and $\rho_i = 0$ otherwise, for which we obtain $z = \varepsilon^2$. Finally, it remains to determine the maximal possible value of $\varepsilon^2$. Since $\epsilon_1$ is of unit norm, its maximal value is attained when we load as much of the normalization into as few of the $\epsilon_{1ii}^2$ as possible. By tracelessness of $\epsilon_1$, at least two of the $\epsilon_{1ii}$ must be nonzero. Thus, $\varepsilon^2$ takes its maximum value of $1/2$ when $\epsilon_1\propto\sigma_3$ in some 2-by-2 block, up to permutation of coordinate labels. That is, a choice of polarizations that maximizes $z$ for $x=y=0$ is
\be
\epsilon_1 =  \tfrac{1}{\sqrt{2}}\sigma_3 \oplus 0_{D-4} \qquad {\rm and} \qquad \epsilon_2 = \tfrac{1}{\sqrt{2}}\sigma_1 \oplus 0_{D-4},
\ee 
which yields the point 
\be
v_1 = \left(0,0,\tfrac{1}{2}\right).\label{eq:v1}
\ee

Let us henceforth consider the case where $z=0$ and explore in $x,y$. This means that the (real, symmetric) matrices $\epsilon_{1,2}$ commute and so are simultaneously diagonalizable. Taking $x=1$, we can ask how large $y$ can be, which will give a vertex of $S$. Since $\epsilon_1 = \pm \epsilon_2$ for $x=1$, we have $y=\Tr(\epsilon_1\cdot\epsilon_1 \cdot \epsilon_1 \cdot \epsilon_1)$. That is, $y$ has positive first and second derivatives in each of the $|\epsilon_{1ii}|$ values; $y$ is therefore maximized when one of the $|\epsilon_{1ii}|$ is as large as possible and the others are equal and small. (If the smaller numbers in the list were unequal, we could always make $y$ larger by shifting some weight back to the element in the list with the largest absolute value.) That is, a choice of polarizations maximizing $y$ for $x=1$ and $z=0$ is
\be
\epsilon_1 = \epsilon_2  = \tfrac{1}{\sqrt{(D-2)(D-3)}}{\rm diag}(1,1,\ldots,-(D-3)), 
\ee
which corresponds to the vertex
\be
v_2 = \left(1,1-\tfrac{3}{D-2}+\tfrac{1}{D-3},0\right).\label{eq:v2}
\ee

Next, still taking $z=0$, we consider a different extreme, setting $x=0$ and maximizing $y$. Again simultaneously diagonalizing $\epsilon_1$ and $\epsilon_2$, we have $y = \sum_i \epsilon_{1ii}^2 \epsilon_{2ii}^2$.  Analogously with the case of $v_1$, we can write $\rho_i = \epsilon_{1ii}^2$ and consider $y$ to be a weighted average over the $\epsilon_{2ii}^2$. Let $\max_i \epsilon_{2ii}^2 \equiv \varepsilon^2$ and, without loss of generality, suppose that $\epsilon_{2ii}^2 = \varepsilon^2$ for $i \in \{1,...,N\}$ for some $N$. Then $y$ is maximized if we take $\rho_i = 1/N$ for $i\in \{1,..., N\}$ and $\rho_i = 0$ otherwise, in which case $y=\varepsilon^2$. Now, by the unit normalization of $\epsilon_2$, $\varepsilon^2$ is maximized when as much of the normalization as possible is loaded into as few terms as possible, {\it i.e.}, $N$ is minimized. Since $\epsilon_1$ is traceless, at least two of the $\rho_i$ are nonzero, so $N\geq 2$. The maximum value of $\varepsilon^2$ thus occurs when $N=2$, which fixes $\epsilon_1 \propto \sigma_3 \oplus 0_{D-4}$. We now must maximize the common absolute value of the first two entries in $\epsilon_{2ii}$, subject to the constraints that $\sum_i \epsilon_{2ii} = 0$, $\sum_i \epsilon_{2ii}^2 = 1$, and, since $x=0$, $\sum_i \epsilon_{1ii} \epsilon_{2ii} = 0$. This last constraint implies that the first two entries in $\epsilon_{2ii}$ have the same sign. Thus, $y$ is maximized for $x=z=0$ for the choice of polarizations
\be 
\epsilon_1 = \tfrac{1}{\sqrt{2}}\sigma_3\oplus 0_{D-4} \qquad {\rm and} \qquad \epsilon_2 = \sqrt{\tfrac{2}{(D-2)(D-4)}}{\rm diag}\left(\tfrac{D-4}{2},\tfrac{D-4}{2},-1,\ldots,-1\right),\label{eq:polarizationv3}
\ee
for which we find the vertex
\be
v_3 = \left(0,\tfrac{D-4}{2(D-2)},0\right).\label{eq:v3}
\ee
Note that the polarization configuration in \Eq{eq:polarizationv3}, and hence the vertex in \Eq{eq:v3}, requires $D\geq 5$.

On the other hand, we can minimize $y/x$ for $z=0$. Using symmetry and reality to diagonalize $H_+ = \mathrm{diag}\,{\vec{h}_+}$, we have $x = |\vec{h}_+ \cdot \vec n|^2$, where $\vec{n} = (1,1,\ldots,1)$, so the Cauchy-Schwarz inequality implies 
\be
x \leq |\vec{h}_+|^2 |\vec{n}|^2 = (D-2) y.\label{eq:xyineq} 
\ee  
\Eq{eq:xyineq} is saturated when $H_+ \propto 1_{D-2}$. If $D$ is even, this choice is possible with
\be
\epsilon_1 = \epsilon_2 = \tfrac{1}{\sqrt{D-2}}{\rm diag}\left(1,-1,1,-1,\ldots\right), 
\ee
which yields
\be
(x,y,z) = \left(1,\tfrac{1}{D-2},0\right).\label{eq:v4even} 
\ee
Let us now consider the odd-dimensional case where $z=0$ and $x=1$. Simultaneously diagonalizing $\epsilon_1$ and $\epsilon_2$, we have $y = \sum_i \epsilon_{1ii}^4$. Again, $y$ has positive first and second derivatives in $|\epsilon_{1ii}|$, so it is minimized when the $|\epsilon_{1ii}|$ are all equal. In odd dimension, this is not possible while retaining tracelessness, so the best one can do, making the $|\epsilon_{1ii}|$ as equal as possible, is the choice 
\be
\epsilon_1 = \epsilon_2 = \sqrt{\tfrac{D-3}{(D-1)(D-2)}}{\rm diag} \Big(\underset{\tfrac{D-1}{2}}{\underbrace{1,\ldots,1}},-\tfrac{D-1}{D-3},...,-\tfrac{D-1}{D-3}\Big),
\ee
which results in the vertex
\be
(x,y,z) = \left(1,\tfrac{2}{D-1} + \tfrac{2}{D-3} - \tfrac{3}{D-2},0\right).\label{eq:v4odd} 
\ee
We can combine \Eqs{eq:v4even}{eq:v4odd} to write the vertex of $S$ as 
\be
v_4 = \left(1,\tfrac{1}{D-2}\left[1+\tfrac{4(D\!\!\!\!\mod 2)}{(D-1)(D-3)}\right],0\right).\label{eq:v4} 
\ee

We note that for both $D=4$ and $D=5$, $v_2$ and $v_4$ are the same point. Moreover, $v_3$ generalizes the third vertex applicable in $D=5$, while the polarization choice for $v_3$ does not apply in $D=4$. In $D\geq 6$, there is one remaining linearly independent vertex, which can be obtained by choosing $\epsilon_1 \cdot \epsilon_2 = 0_{D-2}$, {\it e.g.}, 
\be
\epsilon_1 = \tfrac{1}{\sqrt{2}}\sigma_3 \oplus 0_{D-4} \qquad {\rm and} \qquad \epsilon_2 = \tfrac{1}{\sqrt{2}} 0_{D-4} \oplus \sigma_3,
\ee
which results in the point
\be
v_5 = (0,0,0).\label{eq:v5}
\ee
Together, Eqs.~\eqref{eq:v1}, \eqref{eq:v2}, \eqref{eq:v3}, \eqref{eq:v4}, and \eqref{eq:v5} are the vertices of $\tilde{S}$ given in \Eq{eq:vertices}. They correspond via \Eq{eq:xyzbound} to a set of linearly independent bounds \eqref{eq:D6bounds} that must be satisfied in any gravity theory in $D\geq 6$.

\bibliography{Analyticity_Bounds}
\bibliographystyle{utphys}
\end{document}